\documentclass[aps,prb,twocolumn,amsmath,showpacs]{revtex4}
\usepackage[dvips]{graphicx}
\usepackage{amssymb,amsfonts,amsmath}
\usepackage{color}

\begin{document}

%\author{Vasili Perebeinos and Phaedon Avouris}

%\affiliation{IBM Research Division, T. J. Watson Research Center,
%Yorktown Heights, New York 10598}

\author{Vasili Perebeinos} \email{vperebe@us.ibm.com}
\affiliation{IBM Research Division, T. J. Watson Research Center,
Yorktown Heights, New York 10598}

\author{Phaedon Avouris} \email{avouris@us.ibm.com}
\affiliation{IBM Research Division, T. J. Watson Research Center,
Yorktown Heights, New York 10598}

\title{Inelastic Scattering and Current Saturation in Graphene}
\date{\today}

%\maketitle

\begin{abstract}

We present a study of transport in graphene devices on polar insulating substrates by solving the Bolzmann transport equation in the presence of graphene phonon, surface polar phonon, and Coulomb charged impurity scattering. The value of the saturated velocity shows very weak dependence on the carrier density, the nature of the insulating substrate, and
the low-field mobility, varied by the charged impurity concentration. The saturated velocity of 4 - 8 $\times$ 10$^7$ cm/s calculated at room temperature is significantly larger than reported experimental values. The discrepancy is due to the self-heating effect which lowers substantially the value of the saturated velocity. We predict that by reducing the insulator oxide thickness, which limits the thermal conductance, the saturated currents can be significantly enhanced. We also calculate the surface polar phonon contribution to the low-field mobility as a function of carrier density, temperature, and distance from the substrate.
\end{abstract}

\pacs{72.80.Vp, 72.10.Di, 73.50.Fq}

\maketitle

\section{Introduction}

The excellent transport \cite{Novoselov,Kim,Bolotin1,Du1} and optical properties of graphene \cite{GeimRMP} have attracted strong interest in possible applications of this
material in nanoscale electronics and optoelectronics \cite{GeimScience09,avouris-dev,Xia}. The electrostatic modulation of the graphene channel through gates yields very promising two-dimensional field-effect devices for analog and radio-frequency applications \cite{YuMingRF,ShepardRF}.
Such devices should ideally be operated in the saturation limit \cite{Shepard}. Indeed, it has been shown that the current saturates as the source-drain field is
increased to a few Volts per micron \cite{Shepard,Freitag,Barreiro}. While elastic scattering determines the low-field mobility, the current saturation has been attributed due to the inelastic scattering by either surface polar phonons (SPP) of the polar substrates \cite{Shepard,Freitag} or the intrinsic graphene optical phonons \cite{Barreiro}. In addition to the uncertainty on the nature of the inelastic scattering mechanism, significant heating of the graphene devices operated under high bias conditions is expected. This has recently been measured by Raman spectroscopy \cite{Freitag,Klitzing}. However, little is known about the role of self-heating and elastic scattering on the current saturation.

High bias measurements in graphene\cite{Shepard,Freitag,Barreiro} were analyzed by analogy to 1D carbon nanotubes, where the magnitude of the saturated current is determined by the optical phonon energy responsible for the saturation. However, {\it a priori} an extension of the simple analytical model for the saturated velocity in 2D graphene, as an inverse of the square root of the carrier density, may not be applicable. In Ref. \cite{MacDonald}, using hydrodynamic transport theory including graphene optical phonon scattering, the saturation velocity was found to be weakly carrier density dependent, whereas in Ref. \cite{Guo}, using a Monte Carlo solution of the Boltzmann transport equation (BTE), the saturated velocity was found to follow an inverse square root dependence on the carrier density.

In this work, we explore the effects of Coulomb impurity, graphene phonon, and SPP scattering on different polar substrates on the current saturation in the diffusive transport regime. We find that the self-heating of graphene on SiO$_2$ limits significantly the value of the saturated current. The electronic structure of graphene is described by a $\pi$-orbital tight-binding model with a hoping parameter $t_0=3.1$ eV, which gives a Fermi velocity $v_F=(\sqrt{3}/2)t_0a/\hbar\approx 10^6$ m/s, where $a=0.246$ nm is the graphene lattice constant. For the electron-phonon scattering we use the Su-Schrieffer-Heeger (SSH) model \cite{SSH} to express the  modulation of the $\pi$-orbital overlap $t=t_0-g\delta R_{CC}$ with C-C distance $R_{CC}$. The electron-optical phonon scattering has been calculated by first principles \cite{Piscanec,Louie_eph}. It was found that SSH reproduces fairly well the LDA results for the electron-optical phonon scattering if $g\approx 4.5$ eV/\AA \cite{Piscanec}. On the other hand, electron-electron correlations, taken into account using the GW approximation \cite{LazzeriGW}, give electron-K-point optical phonon coupling corresponding to $g\approx6.5$ eV/\AA \ and coupling to $\Gamma$-point phonons to give $g\approx5.3$ eV/\AA. Therefore, it is expected that electron-phonon coupling in graphene can be renormalized depending on the environment which will screen the electron-electron interactions. In this work we use an average value of $g=5.3$ eV/\AA \ as in Ref.~\cite{PerebeinosCNT1}. The parameters used for SPP scattering on SiO$_2$, HfO$_2$, SiC, and BN polar substrates are given in Table~\ref{tab1}.

\begin{table}[hb]
\caption{\label{tab1} Parameters for the SPP scattering for graphene on SiO$_2$, HfO$_2$, SiC, and hexagonal BN substrates. The surface optical phonon (SO)
energies are obtained from the bulk longitudinal optical (LO) phonons
as  $\hbar\omega_{SO}=\hbar\omega_{LO}\left(\frac{1+1/\epsilon_{0}}{1+1/\epsilon_{\infty}}\right)^{1/2}$.}
\begin{ruledtabular}
\begin{tabular}{ccccc}
 & SiO$_2$\cite{SiO2param} & HfO$_2$\cite{Fischetti} & SiC\cite{SiCbook} & h-BN\cite{hBN}\\
\hline
$\varepsilon_0$ &  3.9 & 22.0  & 9.7 & 5.09\\
$\varepsilon_{i}$ & 3.36 & 6.58 & - & 4.575\\
$\varepsilon_{\infty}$ & 2.40 & 5.03 & 6.5 & 4.10 \\
$\hbar\omega_{SO1}$ in meV & 58.9 & 21.6 & 116.0 & 101.7\\
$\hbar\omega_{SO2}$ in meV & 156.4 & 54.2 & - & 195.7 \\
$F_1^2$ in meV & 0.237 & 0.304 & 0.735 & 0.258\\
$F_2^2$ in meV & 1.612 & 0.293 & - & 0.520 \\
\end{tabular}
\end{ruledtabular}
\end{table}

\section{low-field mobility}

\begin{figure}[h!]
\includegraphics[height=4.50in]{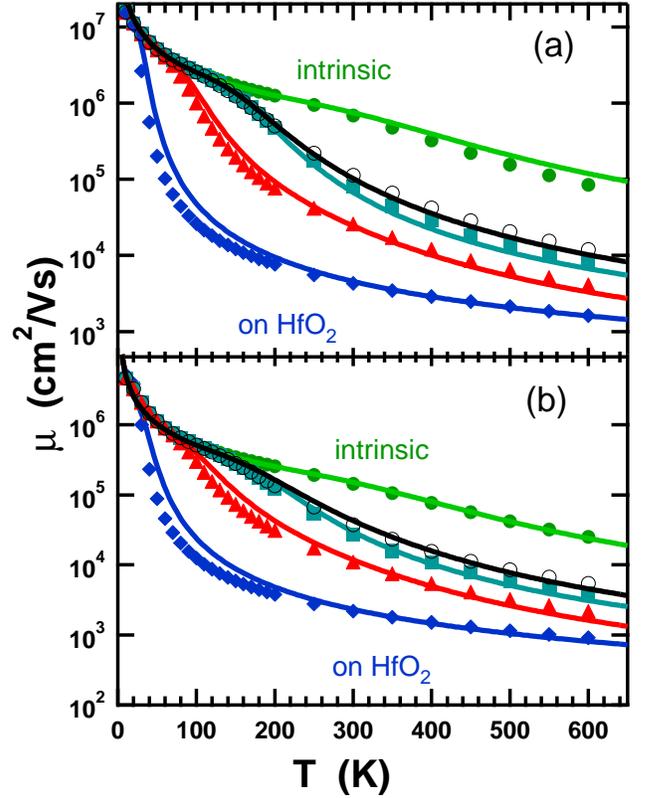}
\caption{\label{Fig1} (Color online) Temperature dependence of the low-field mobility in graphene on different substrates from top to bottom: green circles - intrinsic, black open circles - on BN, cyan squares - on SiC, red triangles - on SiO$_2$, blue open diamonds - on HfO$_2$ at carrier densities (a) $n=10^{12}$ cm$^{-2}$ and (b) $n=5 \times 10^{12}$ cm$^{-2}$. The solid curves are results using Eq. (\protect{\ref{eq10}}) with characteristic sound velocity $v_{ph}=17.3$ km/s, energy of the optical phonon in graphene $\hbar\omega_{op}=0.19$ eV.}
\end{figure}

The low-field mobility in pristine graphene, in the absence of charged impurities and defects, is determined by scattering from the graphene phonons and it is shown in Fig.~\ref{Fig1}. There
are two acoustic phonon branches, transverse (TA) and longitudinal (LA), with an appreciable electron-phonon coupling. Within the SSH model, the TA and LA modes have different angle dependencies for the electron-phonon couplings $\vert M_{k,k+q}\vert^2=D_{ac}^2q^2\hbar/(8NM_C\omega_q)(1\pm\cos(3(\theta_k+\theta_{k+q})))$ \cite{Rice}, where $D_{ac}$ is a deformation potential, $\theta_k$ is a directional angle of wavevector $k$\cite{NoteAngle}, $\omega_q$ is a phonon frequency, $M_C$ is a mass of carbon atom, and $N$ is a number of k-points. Therefore, the acoustic phonon coupling can be approximated by $\vert M_{k,k+q}\vert^2=D_{ac}^2q\hbar/(4NM_Cv_{ph})$, where $v_{ph}$ is a characteristic (LA/TA) sound velocity. The value of deformation potential is given by\cite{Rice}: $D_{ac}=3ga\kappa/(4\sqrt{3})$, where the reduction factor $\kappa=v_{TA}^2/(v_{LA}^2-v_{TA}^2)$ was introduced in Ref.~\cite{Mahan,Andokappa}. For the valence phonon model \cite{PerebeinosTersoff} used here we obtain $D_{ac}\approx 3.7$ eV, consistent with the numerical calculations.

The low-field mobility $\mu$ can be found using Boltzmann theory $\sigma=en\mu=e^2v_F^2 D_n \tau/2$, where $\sigma$ is the conductivity, $n$ is the carrier density, $D_n=2E_F/(\pi\hbar^2 v_F^2)$ is the density of states, $E_F\approx \hbar v_F\sqrt{\pi n}$ is the Fermi energy, and $\tau$ is the scattering time. The latter can be found as:
\begin{eqnarray}
\frac{1}{\tau_k}&=&\frac{2\pi}{\hbar}\sum_q \vert M_{k,k+q}\vert^2 \left(1-\cos(\theta_{k}-\theta_{k+q})\right)
\nonumber \\
&&(N_q\delta(E_k-E_{k+q}+\hbar\omega_q)+
\nonumber \\
&&(N_q+1)\delta(E_k-E_{k+q}-\hbar\omega_q))
\label{eqFermi}
\end{eqnarray}
where $N_{q}$ is the Bose-Einstein phonon occupation number. The summation in Eq.~(\ref{eqFermi}) is replaced by the integral $(1/N)\sum_q=A/(4\pi^2)\int q dq d\theta$ (sum over one spin and one valley), where $A=\sqrt{3}a^2/2$ is the area of the two atom unit cell. The low-field mobility and the scattering rate in the high temperature $T$ limit are given by:
\begin{eqnarray}
\mu_{ac}&=&\frac{e\rho_m\hbar v_F^2v_{ph}^2}{\pi D_{ac}^2}\frac{1}{nk_BT}
\nonumber \\
\frac{1}{\tau_k}&=&\frac{1}{\hbar^3}\frac{E_k}{v_F^2}\frac{D_{ac}^2}{\rho_m v_{ph}^2}k_B T
\label{eq7}
\end{eqnarray}
where $\rho_m=2M_C/A\approx 7.66 \times 10^{-11}$ kg/cm$^{-2}$ is the graphene mass density.
Note a numerical difference of a factor of $4$ in Eq.~(\ref{eq7}) for $\tau$ used in the literature\cite{Rice,DasSarma}. This discrepancy leads, in part, to the large range of deformation potential values from $D_{ac}$=10 eV to 30 eV quoted in the literature \cite{Sugihara,Fuhrer_sio2,Hong,Kim_sus,Ono}. In addition, uncertainty in the sound velocity can also contribute to the spread of the values of the deformation potentials. For example, the deformation potential extracted from the resistivity temperature dependence in graphene on SiO$_2$ at $n=10^{12}$ cm$^{-2}$ in Ref.~\cite{Fuhrer_sio2} would give $D_{ac}=7.1$ eV, if Eq.~(\ref{eq7}) is used with $v_{ph}=17.3$ km/s \cite{PerebeinosTersoff}. The measurements for the electron branch in suspended graphene at  $n=2\times 10^{11}$ cm$^{-2}$ in Ref.~\cite{Kim_sus}  would correspondingly give $D_{ac}=12$ eV. When quantifying the acoustic scattering in different studies it would, therefore, help to report not only the value of $D_{ac}$, but also that of the low-field conductivity extrapolated to room temperature. In our model, it corresponds to $\sigma_{ac}({\rm T=300} K)\approx 0.1$ S. Note that in low mobility samples, due to charged sites in the substrate, an additional temperature dependence from Coulomb scattering may arise \cite{HwangSarmaCoul}.

\begin{figure}[h!]
\includegraphics[height=3.0in]{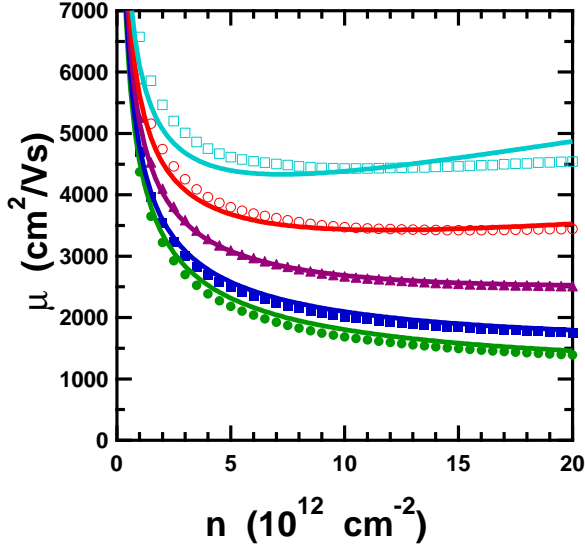}
\caption{\label{Fig1_hfo2} (Color online) Carrier density of the low-field mobility in graphene on HfO$_2$ at T=300 K for distances from the substrate (from top to bottom) $z_0=12.5$ \AA  \ (cyan open squares), $10.0$ \AA \ (red open circles), $7.5$ \AA  \ (magenta triangles), $5.0$ \AA \ (blue squares),  $3.5$ \AA \ (green circles). The solid curves are results using Eq. (\protect{\ref{eq9}}).}
\end{figure}

The two optical phonons at the $\Gamma$ point have couplings $\vert M_{k,k+q}^{s,s'}\vert^2=D_{\Gamma}^2\hbar/(2NM_C\omega_{\Gamma})(1\pm ss'\cos(\theta_k-\theta_{k+q}))$ for LO (- sign) and TO (+ sign) modes respectively \cite{Piscanec,Ando1}, where $D_{\Gamma}=3g/\sqrt{2}\approx 11.2$ eV/\AA \ \cite{Ando2}, $s=1$ for electrons and $s=-1$ for holes. The K-point TO phonon mode has an electron-phonon coupling twice as large \cite{PerebeinosCNT1,Piscanec}, with the angle dependence given by
$\vert M_{k,k+q}^{s,s'}\vert^2=D_{\Gamma}^2\hbar/(NM_C\omega_{K})(1-ss'\cos(\theta_k-\theta_{k+q}))$ \cite{Piscanec}. The effect of the optical phonons (both at $\Gamma$ and $K$) on the low-field mobility can be calculated according to \cite{Ferry}:
\begin{eqnarray}
\mu_{op}=\frac{e\rho_m v_F^2 \omega_{op}}{2\pi D_{op}^2}\frac{1}{nN_{op}}
\label{eq8}
\end{eqnarray}
where $D_{op}=2D_{\Gamma}=22.4$ eV/\AA  \  is an effective electron-optical phonon coupling. The angle integration in Eq.~(\ref{eqFermi}) for $K$-point phonons gives a factor of $3/2$ larger contribution than that for $\Gamma$-point phonons scattering.

The SPP phonons on polar substrates produce
an electric field that couples to the electrons on the nearby graphene.
While the SPP scattering is of lesser
importance in conventional Metal–Oxide–Semiconductor Field-Effect Transistors (MOSFETs)\cite{Fischetti}, it is much more prominent in graphene and carbon nanotubes due to the much smaller vertical dimension of the devices given by the van der Waals distance. It has been invoked
to explain transport under both low and high bias conditions in graphene \cite{Fuhrer_sio2,Shepard,Gunea,Jena} and carbon nanotubes \cite{PerebeinosSPP} on polar substrates. In graphene the SPP coupling is given by \cite{Gunea,Jena}:
\begin{eqnarray}
&&\vert <\Psi^s_k\vert V_{spp}\vert\Psi^{s'}_{k+q}>\vert^2=
\nonumber \\
&=&\frac{1+ss'\cos(\theta_k-\theta_{k+q})}{2}\frac{4\pi^2 e^2F_{\nu}^2}{NAq}e^{-2qz_0}
\label{eq6}
\end{eqnarray}
where $z_0\approx 3.5$  \AA \ is the van der Waals distance between the polar substrate and the graphene flake. The magnitude of the polarization field is
given by the Fr${\rm \ddot{o}}$hlich coupling\cite{MahanSPP}:
$F^2_{\nu}=\frac{\hbar\omega_{SO,\nu}}{2\pi}\left(\frac{1}{\varepsilon_{\infty}+\varepsilon_{{\rm env}}}-
\frac{1}{\varepsilon_{0}+\varepsilon_{{\rm env}}}\right)$,
where $\hbar\omega_{SO,\nu}$ is a surface phonon energy and
$\varepsilon_0$ and $\varepsilon_{\infty}$ are the low- and high-frequency
dielectric constants of the polar substrate, see Table~\ref{tab1}. The screening of the Coulomb interaction by the environment above the polar dielectric is taken into account by $\varepsilon_{{\rm env}}$. Since the screening of the electric field perpendicular to the graphene plane is weak \cite{Louie_eps}, we take $\varepsilon_{{\rm env}}=1$. When there are several
SPP modes with an appreciable coupling, then the low- and high-
frequency $\varepsilon$ are understood as intermediate dielectric functions at
$\omega_i\ll\omega_{SO,\nu}$ for $\epsilon_{0}$ and at $\omega_i\gg\omega_{SO,\nu}$ for
$\varepsilon_{\infty}$ \cite{Fischetti}.

\begin{figure}[h!]
\includegraphics[height=4.0in]{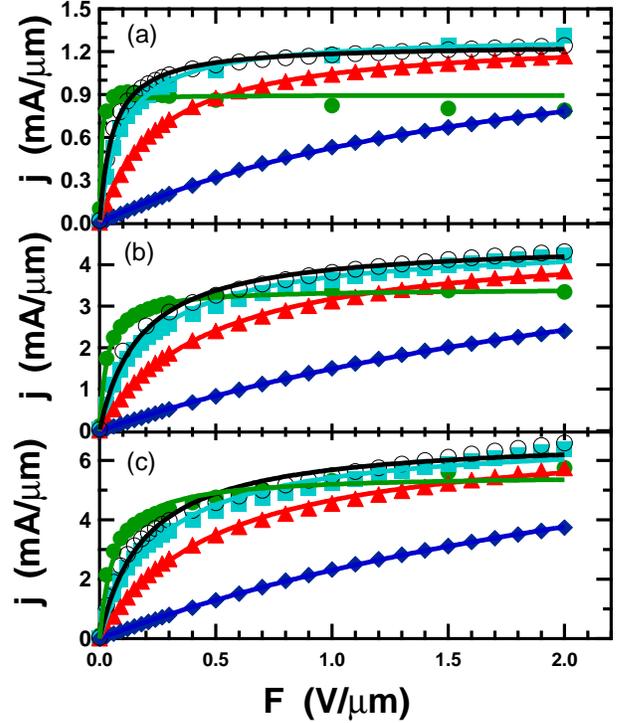}
\caption{\label{Fig2_300} (Color online) Current density electric field dependence in graphene on different polar substrates: green circles - intrinsic, black open circles - on BN, cyan squares - on SiC, red triangles - on SiO$_2$, blue diamonds - on HfO$_2$ at carrier densities (a) $n=10^{12}$ cm$^{-2}$, (b) $n=5 \times 10^{12}$ cm$^{-2}$, and (c) $n=10^{13}$ cm$^{-2}$ and a lattice temperature $T_{amb}=300 $ K. The solid curves are fits to Eq.~(\protect{\ref{eq_satur}}).}
\end{figure}

As shown in Fig.~\ref{Fig1}, the SPP contribution to the low-field mobility can be well approximated by:
\begin{eqnarray}
\mu_{spp, \nu}\approx \sqrt{\frac{\beta}{ \hbar \omega_{\nu}}} \frac{\hbar v_F}{e^2}\frac{ev_F}{F_{\nu}^2}\frac{\exp{\left(k_0z_0\right)}}{N_{spp, \nu}\sqrt{ n}}
\label{eq9}
\end{eqnarray}
which is  a non-monotonic function of carrier density $n$. Here $k_0\approx\sqrt{(2\omega_{SO,\nu}/v_F)^2+\alpha n}$, where parameters $\alpha\approx10.5$ and $\beta\approx 0.153 \times 10^{-4}$ eV are determined to give an overall agreement of the mobility dependence on the carrier density $n$ and distance $z_0$, as shown in Fig.~\ref{Fig1_hfo2}. SiC grown graphene has an intermediate dead layer \cite{SiCdead}, which increases the effective distance $z_0$ and also changes the environmental screening $\varepsilon_{env}$. Both effects, have to be taken into account when direct comparison is made with the experiment. Note that the calculated temperature dependencies shown in Fig.~\ref{Fig1} deviate from the scattering rate temperature dependence given by the SPP phonon occupation number  $N_{spp, \nu}$. An additional temperature dependence arises from the thermal averaging of the scattering rate with the carrier distribution function. We have chosen a set of parameters $\alpha$ and $\beta$ to agree best with the mobility at room temperature.

The calculated low-field mobility from the BTE solution in Fig.~\ref{Fig1} can be well described using Matthiessen's rule:
\begin{eqnarray}
\mu^{-1}=\mu_{ac}^{-1}+\mu_{op}^{-1}+\sum_{nu}\mu_{spp,\nu}^{-1}
\label{eq10}
\end{eqnarray}
where mobility contributions due to the acoustic, optical, and SPP phonons are given by Eq.~(\ref{eq7}), (\ref{eq8}), and (\ref{eq9}), respectively.

\begin{figure}[h!]
\includegraphics[height=3.5in]{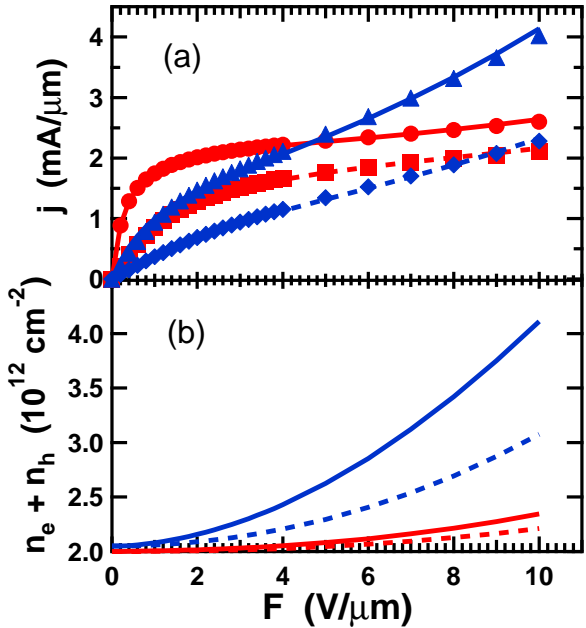}
\caption{\label{Fig3_Coul} (Color online) (a) Current saturation in graphene on SiO$_2$ at different temperatures $T$ in K and Coulomb impurity concentration $n_i$ in  $10^{12}$ cm$^{-2}$: ($T,n_i$)= (300, 0.0) - red circles, (300, 1.35) - red squares, (600, 0.0) - blue triangles, (600, 2.8) - blue diamonds. The solid curves are the
best fits to Eq.~(\ref{eq_satur}) and the corresponding saturated velocities are $v_{sat}=$ 7.26, 7.07, 7.15, 6.91 $\times 10^7$ cm/s, which are the same values within 5\%,  despite the fact that the low-field mobilities vary by a factor of $15$. (b) calculated total carrier density as a function of the electric field for the same cases as in (a). The charge carrier density was fixed at $n=n_e-n_h$= 2 $\times 10^{12}$ cm$^{-2}$.}
\end{figure}

\section{Current Saturation}

At high bias, the transport in graphene is usually described by the saturated current model \cite{Shepard,PerebeinosCNT1,Pierret}:

\begin{eqnarray}
j=e\eta\frac{\mu F}{1+\mu F/v_{sat}}
\label{eq_satur}
\end{eqnarray}
where $\eta=n_e+n_h$ is the total carrier density, $v_{sat}$ is a saturation velocity, and $F$ is an electric field. Note that in the model $\mu_e=\mu_h$. Within simple model, in the full saturation regime only carriers around the Fermi energy $E_{F}$ in the energy window $E_F\pm\hbar\Omega/2$ contribute to the current. Here $\Omega$ is the characteristic frequency of the phonon responsible for the current saturation. The saturated velocity can then be calculated and for $E_F>\hbar\Omega/2$:
\begin{eqnarray}
v_{sat}\approx\frac{2}{\pi}\Omega\frac{\hbar v_F}{E_F}=\frac{2}{\pi}\frac{\Omega}{\sqrt{\pi n}}
\label{eq12}
\end{eqnarray}
Note again a difference in the numerical prefactor used in the literature\cite{Shepard,Freitag,Barreiro} $1$ versus $2/\pi$ which affects the numerical value of the characteristic phonon energy $\hbar\Omega$ extracted from the experiments \cite{Shepard,Freitag}.

\begin{figure}[h!]
\includegraphics[height=3.0in]{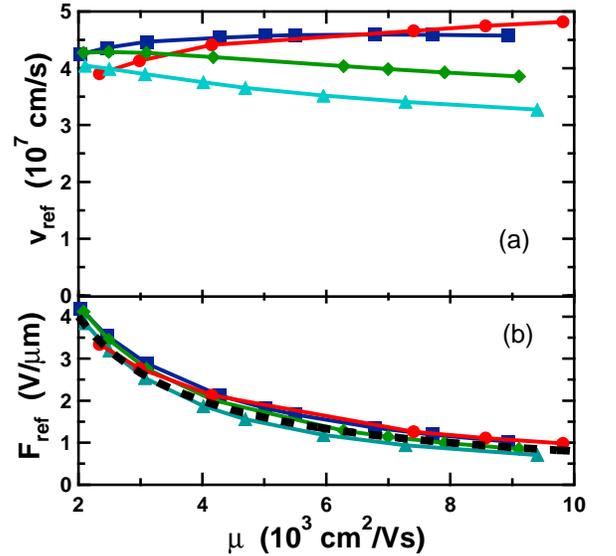}
\caption{\label{Fig_scale} (Color online) (a) Reference velocity $v_{ref} \approx v_{sat}/2$ and (b) reference field $F_{{ ref}}\approx v_{sat}/\mu$ dependencies on the low-field mobility. The results are obtained from the calculations at room temperature for graphene on SiO$_2$ for different impurity concentrations, which determine the low-field mobility\cite{DasSarma2}. Calculations are  for fixed charge carrier densities $n$ (in 10$^{12}$ cm$^{-2}$) of 1.0 - red circles, 2.0 - blue squares, 3.0 - green diamonds, 5.0 - cyan triangles. The black dashed curve in (b) shows $F_{{ref}}=v_{sat}/\mu$ dependence with $v_{sat}=8\times 10^7$ cm/s.}
\end{figure}

The current densities for graphene on different substrates as a function of electric field are shown in Fig.~\ref{Fig2_300}. When phonons are kept in thermal equilibrium at $T_{amb}=300$ K, the current does not show full saturation for the experimentally relevant source-drain fields up to 2 V/$\mu$m. In Fig.~\ref{Fig2_300}a, the current shows negative differential conductance for scattering by intrinsic  graphene phonons. This behavior is due to the deviation of the bandstructure from the linear band dispersion, similar to the effect of the non-parabolicity in carbon nanotubes \cite{PerebeinosCNT1}. While the current at high bias in Fig.~\ref{Fig2_300} is found to be larger for graphene on polar insulators with larger SPP phonon energy, the saturated velocity as obtained from the fit to Eq.~(\ref{eq_satur}) does not obey Eq.~(\ref{eq12}).

\begin{figure}[h!]
\includegraphics[height=2.60in]{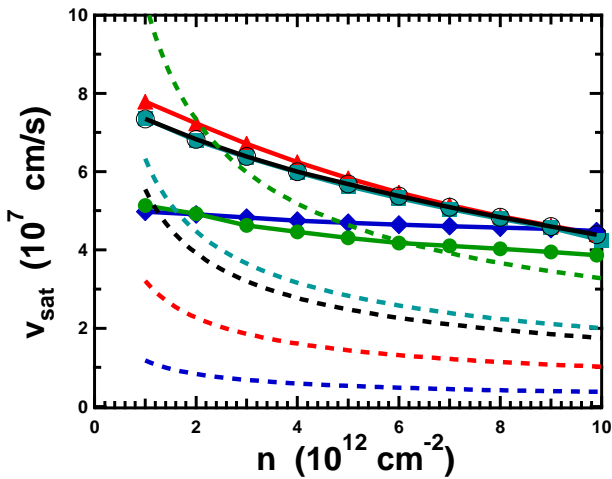}
\caption{\label{Fig_veloc_sat} (Color online) Carrier density dependence of the saturated velocity in graphene on different substrates at room temperature: green circles - intrinsic, black open circles - on BN, cyan squares - on SiC, red triangles - on SiO$_2$, blue diamonds - on HfO$_2$. The dashed curves are predictions using Eq. (\protect{\ref{eq12}}) with optical phonon energy $\hbar\omega_{SO1}$ from Table~\protect{\ref{tab1}}, from top to bottom: green - intrinsic, cyan - on SiC, black - on BN, red - on SiO$_2$, blue - on HfO$_2$.}
\end{figure}

First we explore the dependence of the saturated velocity on the low-field mobility. The latter is strongly affected by the quality of the substrate, which introduces Coulomb impurity scattering \cite{Cheanov,AndoCoulomb,MacDonaldCoulomb,DasSarma2,Fuhrer,HwangSarmaCoul,Fuhrer_sio2,Wendy}, resonant scatterers \cite{Ostrovky1,Katsnelson2,Stauber}, electron-hole puddles \cite{puddle}, and surface roughness \cite{Katsnelson1,Heinz_rough}. In Fig.~\ref{Fig3_Coul}a we show current-voltage characteristics of graphene on SiO$_2$ with a variable Coulomb impurity concentration $n_i$ following Ref.~\cite{DasSarma2}. Despite the fact that the low-field mobilities vary by a factor of $15$, the saturated velocity, as obtained form the fit to Eq.~(\ref{eq_satur}), is essentially the same. The electronic temperature increases with the electric field, such that minority carriers contribution to the current becomes significant. This effect is more prounced at elevated lattice temperatures as shown in Fig.~\ref{Fig3_Coul}b.
%Note that as the bias increases the carrier density increases as well, especially at elevated lattice temperatures, as %shown in Fig.~\ref{Fig3_Coul}b.

The systematic behavior of the low-field mobility, which is modified by the impurity concentration, on the saturated velocity is shown in Fig.~\ref{Fig_scale}. Since the current does not fully saturate even at large fields, we have chosen to plot in Fig.~\ref{Fig_scale} a reference velocity  $v_{{ref}}$ calculated at the reference field $F_{{ref}}$ where the mobility is reduced by a factor of two. If Eq.~(\ref{eq_satur}) is applicable, then the reference velocity is equal to half the saturated velocity $v_{{sat}}$ and is reached at a reference field of $F_{{ref}}=v_{sat}/\mu$. We find that the reference velocity is about $3 - 5 \times 10^{7}$ cm/s and has little dependence on either carrier density or  low-field mobility in the parameter space considered here. The reference field is inversely proportional to the low-field mobility as shown in Fig.~\ref{Fig_scale}b, which justifies the functional form Eq.~(\ref{eq_satur}) used to analyze the BTE results.

\section{Self-Heating effect on current saturation}

The analysis of both experimental\cite{Shepard,Freitag,Barreiro} and simulation \cite{Guo} results had often relied on Eq.~(\ref{eq12}). In Fig.~\ref{Fig_veloc_sat} we show that Eq.~(\ref{eq12}) fails qualitatively to describe the results of the BTE simulations. In particular, saturated velocity in graphene on SiO$_2$ is predicted to be very similar to that in graphene on BN and SiC substrates, although SPP phonon energies in the latter are almost a factor of two larger. While Eq.~(\ref{eq12}) predicts a variation of $v_{sat}$ by more than an order of magnitude for the range of $n$ and $\hbar\Omega$ used in Fig.~\ref{Fig_veloc_sat}, we find the values of $v_{sat}$ from the fits to Eq.~(\ref{eq_satur}) to be within $4 - 8 \times 10^{7}$ cm/s. The saturated velocity in intrinsic graphene shown by green curve in Fig.~\ref{Fig_veloc_sat} agrees well with the Monte Carlo BTE solution in low density limit in Ref.~\cite{Goldman}.
At the same time, the reported experimental values\cite{Shepard,Freitag} of $v_{sat}$ at densities of $10^{13}$ cm$^{-2}$ are below $10^7$ cm/s, which is at least a factor of $4$ smaller than that in Fig.~\ref{Fig_veloc_sat}.

We suggest that one of the factors for the apparent discrepancy is the self-heating effect. The temperature rise was shown\cite{Freitag} to be proportional to the Joule losses $T=T_{amb}+jF/r$, where $T_{amb}$ is the ambient temperature and the thermal conductance $r$ controls the heat dissipation. This simple picture applies only for graphene regions away from the contacts such that both the thermal contact resistance and the substrate thermal conductivity determine the value of $r$. Note that the upper bound for $r$, which corresponds to zero contact thermal resistance, is determined by the thermal conductivity and thickness of the insulating substrate. For example, thermal conductivity of SiO$_2$ is $\kappa\approx 1.4$ W/(mK) \cite{CRC} and for the insulator height of $h=300$ nm a maximum value of $r=\kappa/h\approx0.47$ kW/(K cm$^2$) is expected. However, this upper bound can be significantly reduced due to the thermal contact resistance as in Ref.~\cite{Freitag}.

\begin{figure}[h!]
\includegraphics[height=4.0in]{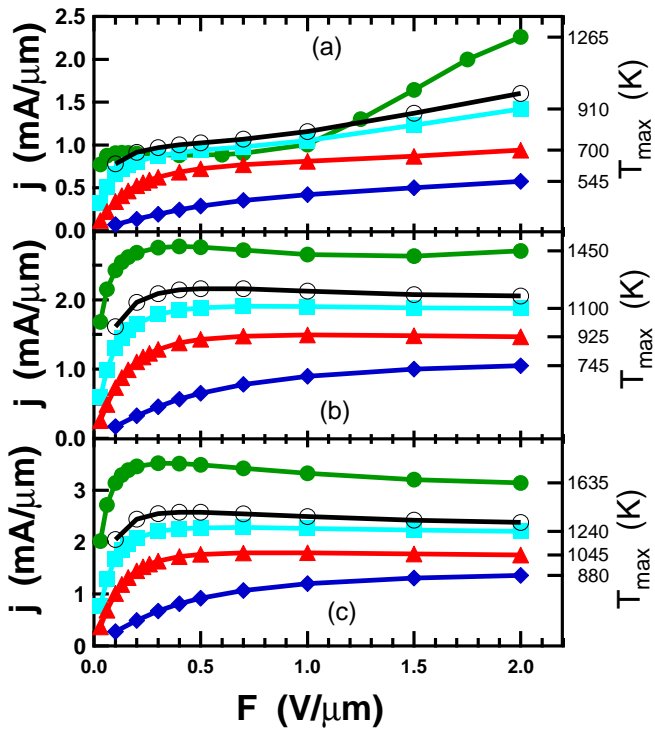}
\caption{\label{Fig2_self} (Color online) Current density electric field dependence in graphene on different polar substrates: green circles - intrinsic, black open circles - on BN, cyan squares - on SiC, red triangles - on SiO$_2$, blue diamonds - on HfO$_2$ at carrier densities (a) $n=10^{12}$ cm$^{-2}$, (b) $n=5 \times 10^{12}$ cm$^{-2}$, and (c) $n=10^{13}$ cm$^{-2}$ at self consistent lattice temperatures $T=T_{amb}+j(T)F/r$ for $r=0.47$ kW/(K cm$^2$), corresponding to the ideal thermal contact resistance with 300 nm SiO$_2$. The vertical axis on the right shows corresponding values of the maximum temperatures at 2 V/$\mu$m.}
\end{figure}

In the presence of the SPP scattering the role of the thermal contact resistance is minimized because electrons can give some of their energy directly to the substrate SPP phonons \cite{Rotkin,MahanKapitza}. In our ``self-heating model'' we   assume that SPP and graphene phonons are heated to the same temperature, which is proportional to the Joule losses found self-consistently, i.e. $j=j(T)$. In substrates with much higher thermal conductivities, such as SiC, BN and HfO$_2$, the thermal contact resistance would dominate the value of $r$. In this case a full self-consistent solution including thermal transport in the substrate, which determines the SPP phonon temperature, would be needed. In Fig.~\ref{Fig2_self} we find that the current densities drop by up to a factor of four at high biases as a result of self-heating for $r=0.47$ kW/(Kcm$^2$). Moreover, the current now shows true saturation at experimental source-drain fields, which are much smaller than $F_{\rm ref}$ calculated in Fig.~\ref{Fig_scale}b at room temperatures, and the high bias currents are comparable to those reported in Ref.~\cite{Shepard,Freitag}.

\begin{figure}[h!]
\includegraphics[height=3.20in]{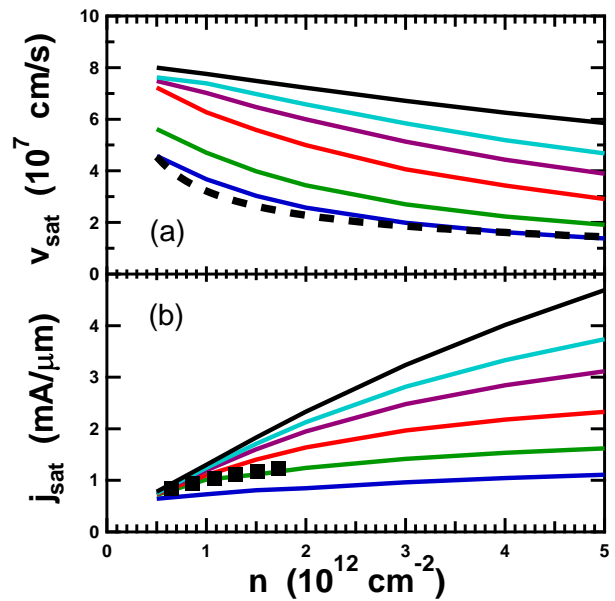}
\caption{\label{Fig_veloc_bach} (Color online) Carrier density dependence of (a) the saturated velocity and (b) saturated current in graphene on SiO$_2$ using the self-heating model at different thermal conductances $r$ in kW/(K cm$^2$) from top to bottom: black - T=300 K or $r=\infty$, cyan - $r=4.0$, magenta - $r=2.0$, red - $r=1.0$, green - $r=0.5$, blue - $r=0.25$, which correspond to the ideal thermal resistance to SiO$_2$ with an effective thickness $h=\kappa/r=$ 0, 35, 70, 140, 280, and 560 nm correspondingly. The dashed black curve in (a) is calculated using Eq.~(\protect{\ref{eq12}}) for SPP phonon in SiO$_2$. The solid squares in (b) show fitted saturated currents to Eq.~(\ref{eq_satur}) from Ref.~\cite{Barreiro}.}
\end{figure}

At high density, the self-heating involving the intrinsic graphene phonons is predicted here to lead to negative differential conductance. This effect has served as an experimental signature of the self-heating in suspended carbon nanotubes \cite{Dai}. In the presence of SPP scattering the negative differential conductance is less pronounced and the self-heating effect leads to current saturation in Fig.~\ref{Fig2_self}. The different outcome of self-heating effects leading to either negative differential conductance or current saturation can well be understood by the difference in the graphene phonon and SPP phonon energies. Indeed, the current density in the diffusive regime is $j\propto F\tau$. According to Eq.~(\ref{eqFermi}), the self-heating reduces the scattering time as $\tau\propto (1+2N_{op})^{-1}$, such that $j\propto F/(1+2N_{op})$. In the high temperature limit $N_{op}\propto k_BT/\hbar\omega_{op}$ and current saturates as $j\propto F/(k_BT_{amb}+jF/r)$, where we have used the self-heating temperature $k_BT=k_BT_{amb}+jF/r$. The saturation current in this simple picture depends on $r$ as $j_{sat}\propto r^{0.5}$, which is in qualitative agreement with a full BTE solution shown in Fig.~\ref{Fig_veloc_bach}b.
On the other hand, the large intrinsic graphene optical phonon leads to an activated temperature dependence of the scattering time, which produces the negative differential conductance in Fig.~\ref{Fig2_self}.

\begin{figure}[h!]
\includegraphics[height=3.60in]{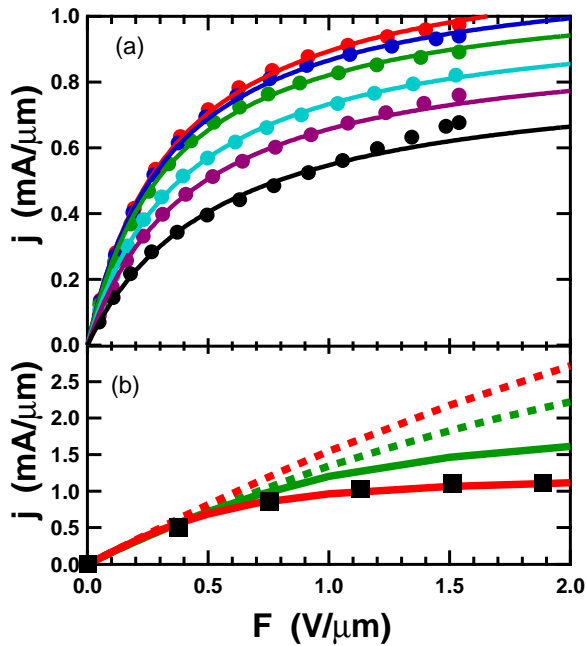}
\caption{\label{Fig3} (Color online) (a) Circles show data points from Ref.~\cite{Barreiro} for $V_g=$24, 21, 18, 15, 12, 9 V from top to bottom. The solid curves are fits to Eq.~(\ref{eq_satur}) with $\eta=0.072 \times V_g$ 10$^{12}$ cm$^{-2}$, where $V_g$ is in V. The fit results are shown in Fig.~\ref{Fig_veloc_bach}b. (b)
Modeling of the current-voltage characteristics measured in Ref.~\cite{Freitag}. The green and red dashed curves, correspondingly, are BTE solutions for graphene phonon scattering and both graphene and SPP phonon scattering at $T_{amb}=300$ K including Coulomb scattering (see text). The green and red solid curves are self-heating model calculations with $r=0.278$ kW/(K cm$^2$)~\cite{Freitag} including graphene phonon only, and both SPP and graphene phonon scattering, respectively. Black squares are experimental results from Ref.~\cite{Freitag}.}
\end{figure}

The effect of the thermal conductance $r$ on velocity saturation is explored in Fig.~\ref{Fig_veloc_bach}. As $r$ is reduced from infinity to 0.25 kW/(K cm$^2$), the saturated velocity drops by a factor of four. For small values of $r$ and large densities, we find that Eq.~(\ref{eq_satur}) fails to fit the results of calculations in the full range of fields and we have chosen to fit a range of fields from 0.5 to 1 V/$\mu$m, where saturation is almost attained. The calculated saturated current in Fig.~\ref{Fig_veloc_bach}b for $r=0.5$ kW/(K cm$^2$) agrees remarkably well with recent measurements in Ref.~\cite{Barreiro}. The data from Ref.~\cite{Barreiro} are fitted to Eq.~(\ref{eq_satur}) as shown in Fig.~\ref{Fig3}a. Note that accidentally Eq.~(\ref{eq12}) reproduces fairly well the results of self-heating model for $r=0.25$ kW/(K cm$^2$) in Fig.~\ref{Fig_veloc_bach}a.

In the experiment described in Ref.~\cite{Freitag}, the thermal conductance $r$ was directly measured and does not have to be a fit parameter in modeling the experimental I-V curve. The observed mobility of
about 1000 cm$^2$/Vs in Ref.~\cite{Freitag} can be reproduced in our calculations by assuming scattering with charged impurities~\cite{DasSarma2} of density $n_i=4.5 \times 10^{12}$ cm$^{-2}$ and with a smaller $n_i=3.5 \times 10^{12}$ cm$^{-2}$ in the presence of SPP scattering from the  SiO$_2$ substrate. The charge carrier density was fixed by the gate voltage at $n\approx 10^{13}$ cm$^{-2}$ in Ref.~\cite{Freitag}. In Fig.~\ref{Fig3}b we show that the calculated current is significantly larger at high biases in the presence of both intrinsic graphene and SPP phonon scattering at room temperatures. Most importantly, the current does not show the saturation that is observed in the experiment. On the other hand, using the experimentally measured temperatures, our self-heating model with $r=0.278$ kW/(K cm$^2$) including the SPP scattering very nicely reproduces the experiment. However, if we assume self-heating model including  only the graphene  phonons (no SPP) we do not find full saturation, even at fields up to 2 V/$\mu$m, and the calculation overestimates the measured currents at high biases.

\section{conclusion}

In conclusion, we explored the effect of current and velocity saturation on the carrier density and impurity concentration including intrinsic graphene phonons and SPP scattering on polar substrates. The observed full current saturation is simulated only when we account for the self-heating. Without self-heating, the current densities are predicted to be too high for either graphene phonon scattering or SPP scattering. While impurity scattering modifies substantially the low-field mobility, it has little effect on the saturated velocity. Furthermore, the saturated velocity depends very weakly on carrier density and the choice of substrate. These dependencies served as the basis for invoking SPP scattering for graphene on SiO$_2$ as the velocity saturation mechanism in Ref.~\cite{Shepard,Freitag}. A more direct experimental confirmation for the SPP role in velocity saturation is desirable.  We predict a factor of four  enhancement  of the saturation current if self-heating effects are minimized. This can be achieved by using an appropriate choice of substrate with high thermal conductivity, scaled down insulator thickness, and keeping the graphene/substrate contact thermal resistance low.

\acknowledgments{}

We gratefully acknowledge insightful discussions with Francesco Mauri and the co-authors of Ref.~\cite{Barreiro}, Aniruddha Konar (University of Notre Dame), and Inanc Meric (Columbia University). We thank Marcus Freitag and Amelia Barreiro for providing data for Fig.~\ref{Fig3}.


\begin{thebibliography}{99}
\bibitem{SiO2param} We use some avaraged values from several reports \cite{Fischetti,Gunea,Jena}.
\bibitem{NoteAngle} For wavevector $k$ arbitrary away from $K$-point, the angle $\theta_k$ is given by  $\exp{(i\theta_k)}=\gamma_k/\vert \gamma_k \vert$, where $\gamma_k=\exp{\left(-ik_ya/\sqrt{3}\right)}+2\cos{\left(k_xa/2\right)}\exp{\left(ik_ya/(2\sqrt{3})\right)}$.
\end{thebibliography}
\end{document}